\documentclass[twocolumn,10pt]{IEEEtran}
\usepackage{cite}
\usepackage{citesort}
\usepackage{graphicx,import}
\usepackage{amsmath}
\usepackage{amsfonts}
\usepackage{epstopdf}
\usepackage{xcolor}
\usepackage{amssymb,bm,upgreek}
\usepackage{algorithm}
\usepackage{algorithmic}
\usepackage{multirow}
\usepackage{verbatim}
\usepackage{subfigure}

\usepackage{import}

\DeclareMathOperator\erf{erf}

\let\ss= \scriptscriptstyle
\newcommand{\RX}{\textnormal{RX}}
\newcommand{\R} {\textnormal{RX}}
\newcommand{\TX}{\textnormal{TX}}
\newcommand{\T}{\textnormal{TX}}
\newcommand{\FC} {\textnormal{FC}}
\newcommand{\trans}{\textrm{trans}}
\newcommand{\report}{\textrm{report}}
\newcommand{\metre}{\textnormal{m}}

\newcommand{\m}{\textnormal{m}}

\newcommand{\ob}{\textnormal{ob}}

\newcommand{\D}{{\ss\textnormal{D}}}

\newcommand{\s}{\textnormal{s}}
\newcommand{\n}{\textnormal{n}}

\newcommand{\ad}{\textnormal{ad}}

\newtheorem{theorem}{Theorem}

\begin{document}

\title{Maximum Likelihood Detection for Cooperative Molecular Communication}

\author{\IEEEauthorblockN{Yuting Fang${}^\dag$, Adam Noel${}^\ddag$, Nan Yang${}^\dag$, Andrew W. Eckford${}^\sharp$, and Rodney A. Kennedy${}^\dag$}
\IEEEauthorblockA{${}^\dag$Research School of Engineering, Australian National University, Canberra, ACT, Australia\\}
\IEEEauthorblockA{${}^\ddag$School of Electrical Engineering and Computer Science, University of Ottawa, Ottawa, ON, Canada\\}
\IEEEauthorblockA{${}^\sharp$Department of Electrical Engineering and Computer Science, York University, Toronto, ON, Canada}}



\maketitle

\begin{abstract}
In this paper, symbol-by-symbol maximum likelihood (ML) detection is proposed for a cooperative diffusion-based molecular communication (MC) system. In this system, a fusion center (FC) chooses the transmitter's symbol that is more likely, given the likelihood of the observations from multiple receivers (RXs).
We propose three different ML detection variants according to different constraints on the information available to the FC, which enables us to demonstrate trade-offs in their performance versus the information available.
The system error probability for one variant is derived in closed form. Numerical and simulation results show that the ML detection variants provide lower bounds on the error performance of the simpler cooperative variants and demonstrate that majority rule detection has performance comparable to ML detection when the reporting is noisy. 
\end{abstract}

\IEEEpeerreviewmaketitle

\section{Introduction}\label{sec:intro}

Molecular communication (MC) has been heralded as one of the most promising paradigms to implement communication in bio-inspired nanonetworks, due to the potential benefits of bio\text{-}compatibility and low energy consumption; see \cite{Weisi2016}. In MC, the information transmission between devices is realized through the exchange of molecules; see \cite{Andrew_Book}. The simplest molecular propagation mechanism is free diffusion, where the information-carrying molecules can propagate from a transmitter (TX) to a receiver (RX) via Brownian motion. One of the primary challenges posed by diffusion-based MC is that its reliability rapidly decreases when the TX-RX distance increases.
A naturally-inspired approach, which makes use of the envisioned collaboration between nanomachines, is allowing for multiple RXs to share information for cooperative detection. Often, cells or organisms share the same information to achieve a specific task, e.g., calcium signaling; see \cite{Tadashi2010}.

The majority of the existing MC studies have focused on the modeling of a single-RX MC system. Recent studies, e.g.,~\cite{Einolghozati2016,Koo2016,Andrew2015}, have expanded the single-RX MC system to the multi-RX MC system. However, these studies have not explored the potential of \emph{active cooperation} among multiple RXs in determining the TX's intended symbol sequence in a multi-RX MC system. To address this gap, our contributions in~\cite{GC2016,TMBMC2016,simplified} analyzed the error performance of a cooperative diffusion-based MC system where a fusion center (FC) device combines the binary decisions of distributed RXs to improve the detection of a TX's symbols.

In other fields of communications, e.g., wireless communications, the maximum likelihood (ML) detector is commonly used to optimize detection performance; see \cite[Ch,\;5]{Digital}. In the MC domain, the ML sequence detector has been considered for optimality in several studies. 
For example, \cite{Adam2014,Meng2014} considered variations to modify the Viterbi algorithm and reduce the computational complexity of optimal detection.
However, the high complexity of sequence detection is a significant barrier to implementation in the MC domain, even when applying simplified ML algorithms.

We note that the (suboptimal) symbol-by-symbol ML detector requires less computational complexity, compared to the ML sequence detector. Motivated by this, \cite{Mahfuz2015,Amit2016} considered symbol-by-symbol ML detection for MC, but focused on a single RX only. Recently, \cite{Trang2017,Reza2017} considered cooperative ML detection for MC. However, \cite{Trang2017,Reza2017} considered detection of an event, rather than modulated information from a TX.

In this paper, we present symbol-by-symbol ML detection for the cooperative diffusion-based MC system, i.e., the FC chooses the TX symbol that is more likely, given the (joint) likelihood of its observations from all the individual RXs. The significance of this paper is that our results provide lower bounds on the error performance of the simpler cooperative variants in ~\cite{GC2016,TMBMC2016,simplified}. We demonstrate the good performance of these simpler cooperative variants (which are more likely to be realizable in biological environment) relative to symbol-by-symbol ML variants, particularly when we impose reasonable constraints on the knowledge available at the FC.

In our system, the transmission of each information symbol from the TX to the FC via the RXs is completed in two phases. In the first phase, the TX sends a symbol to all RXs. In the second phase, the RXs send their detected information to the FC. We consider relatively simple RXs and keep the relatively high complexity associated with ML detection at the FC\footnote{This consideration is because that the FC could have a direct interface with the macroscopic world and easier access to computational resources.}. In this work we consider the transmission of a sequence of binary symbols and the resultant intersymbol interference (ISI) in the design and analysis of the cooperative MC system. For convenience, we refer to the FC-estimated previous symbols as local history and the perfect knowledge of the previous symbols as genie-aided history. In our design, the FC chooses the current symbol using the FC's \emph{local history}. To the best knowledge of the authors, our work is the \emph{first} to apply symbol-by-symbol ML detection to a cooperative MC system with multiple communication phases.

We consider two different reporting scenarios according to the FC's knowledge of the observations at the RXs, namely, perfect reporting and noisy reporting. The error performance in perfect reporting provides a lower bound on that in noisy reporting. Also, the perfect reporting scenario is appropriate for non-diffusive reporting channels such as neurons; see \cite{Junichi2017}. In perfect reporting, based on different levels of the knowledge of the observations at the RXs, we propose two ML detection variants, namely, 1) full knowledge ML detection (F-ML) and 2) limited knowledge ML detection (L-ML). In noisy reporting, we consider one detection variant between the RXs and the FC, namely, decode-and-forward (DF) with a single type of molecule and ML detection at the FC (SD-ML).

Our contributions are summarized as follows. We use the genie-aided history to derive the system error probability of L-ML in closed form; that of F-ML and SD-ML will be considered in future work. Using simulation and numerical results, we demonstrate the FC's effectiveness in obtaining local history for all ML detection variants. We corroborate the accuracy of our analytical results. Also, our results reveal trade-offs among the performance, the knowledge of previously-transmitted symbols, and computational complexity.

The rest of this paper is organized as follows. In Section \ref{sec:system model}, we describe the system model and analytical preliminaries for ML detection design and analysis. In Section \ref{sec:ML detection design}, we present the design of three ML detection variants (F-ML, L-ML, and SD-ML) and the error performance analysis of L-ML. Numerical and simulation results are provided in Section \ref{sec:Numerical}. In Section \ref{sec:con}, we conclude this work and identify future directions.

\section{System Model and Preliminaries}\label{sec:system model}

\subsection{System Model}


\begin{figure}[!t]
\centering
\includegraphics[width=0.9\columnwidth]{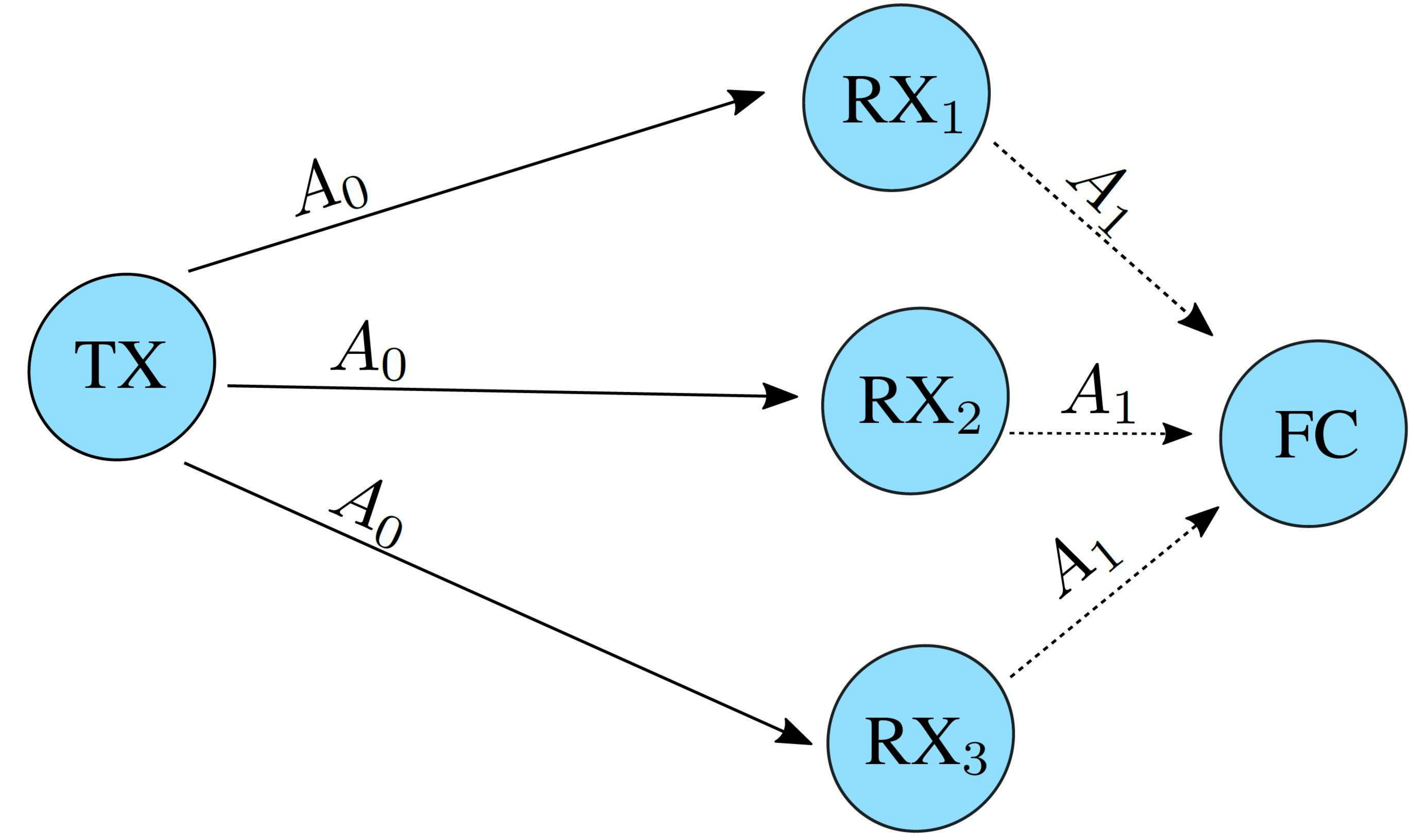}
\caption{An example of a cooperative MC system with $K=3$, where the transmission from the TX to the RXs is represented by solid arrows and the transmission from the RXs to the FC is represented by dashed arrows.}
\label{system model}
\end{figure}

We consider a cooperative MC system in a three-dimensional space based on \cite{GC2016,TMBMC2016,simplified}, which consists of one TX, a ``cluster'' of $K$ RXs, and one device acting as an FC.
An example of the cooperative MC system is illustrated in Fig.~\ref{system model}. For reliable reporting, we generally assume that the FC is close to the cluster of RXs. For tractability, we assume that the system has a symmetric topology, such that the distances between the TX and the RXs are identical and the distances between the RXs and the FC are also identical. We also assume that all RXs and the FC are passive spherical observers such that molecules can diffuse through them without reacting. Accordingly, we denote $V_{\ss\R_k}$ and $r_{\ss\R_k}$ as the volume and the radius of the $k$th RX, $\RX_k$, respectively, where $k\in\{1,2,\ldots,K\}$. We also denote $r_{\ss\FC}$ as the radius of the FC. We use the terms ``sample'' and ``observation'' interchangeably to refer to the number of molecules observed by a RX or the FC at some time $t$. We assume that the time between samples is sufficiently large and the distance between RXs is sufficiently large for all individual observations to be independent of each other.

In this paper, the transmission interval time from the TX to the RXs is denoted by $t_{\trans}$ and the report interval time from the RXs to the FC is denoted by $t_{\report}$. Thus, the symbol interval time from the TX to the FC is given by $T = t_{\trans}+t_{\report}$. We emphasize that the symbol interval time $T$ is the same in the perfect and the noisy reporting scenarios.

In the following, we present the timing schedule of the TX, the RXs, and the FC. At the beginning of the $j$th symbol interval, i.e., $(j-1)T$, the TX transmits $W_{\ss\T}[j]$. The TX transmits $W_{\ss\T}[j]$ to the RXs over the diffusive channel via type $A_0$ molecules which diffuse independently. The TX uses ON/OFF keying to convey information, i.e., the TX releases $S_{0}$ molecules of type $A_0$ to convey information symbol ``1'' with probability $\textrm{Pr}(W_{\ss\T}[j]=1)=P_1$, where $\textrm{Pr}(\cdot)$ denotes probability, but no molecules to convey information symbol ``0''. The TX then keeps silent until the start of the $(j+1)$th symbol interval. We denote $L$ as the number of symbols transmitted by the TX. We define\footnote{Throughout this paper, $W$ is a single information symbol and $\textbf{W}$ is a vector of information symbols.} $\textbf{W}_{{\ss\T}}^{l}=\left\{W_{\ss\T}[1],\ldots,W_{\ss\T}[l]\right\}$ as an $l$-length subsequence of the symbols transmitted by the TX, where $l\leq{L}$.

Each $\RX_{k}$ observes type $A_0$ molecules over the $\TX-\RX_{k}$ link and takes $M_{\ss\RX}$ samples in each symbol interval at the same times. The time of the $m$th sample by each RX in the $j$th symbol interval is given by $t_{\ss\R}(j,m) = (j-1)T + m\Delta{t_{\ss\R}}$, where $\Delta{t_{\ss\R}}$ is the time step between two successive samples by each RX, $m\in\left\{1,2,\ldots,M_{\ss\RX}\right\}$, and $M_{\ss\RX}\Delta{t_{\ss\R}}<t_\trans$. For SD-ML, at the time $(j-1)T + t_{\trans}$, each RX transmits type $A_{1}$  molecules via a diffusion-based channel to the FC. The time of the $\tilde{m}$th sample by the FC in the $j$th symbol interval is given by $t_{\ss\FC}(j,\tilde{m})=(j-1)T+t_{\trans}+\tilde{m}\Delta{t_{\ss\FC}}$, where $\Delta{t_{\ss\FC}}$ is the time step between two successive samples by the FC and $\tilde{m}\in\left\{1,2,\ldots, M_{\ss\FC}\right\}$.

%

\subsection{Preliminaries}
In this subsection, we establish some fundamental preliminary results based on \cite{Adam2014}. These results are needed to evaluate the likelihoods in the ML detection variants in Section \ref{sec:ML detection design}.

\subsubsection{$\TX-\RX_k$ Link}\label{TX-RX link}
We first evaluate the probability $P_{\ob}^{({\ss{\T},{\ss\R_k}})}\left(t\right)$ of observing a given type $A_0$ molecule, emitted from the TX at $t=0$, inside $V_{\ss\R_k}$ at time $t$. Given independent molecule behavior and assuming that the RXs are sufficiently far from the TX, we use \cite[Eq.~(13)]{Adam2014} to write
\begin{align}\label{probability}
P_{\ob}^{({\ss{\T},{\ss\R_k}})}(t) = \frac{V_{\ss\R_k}}{(4\pi D_{0}t)^{3/2}}\exp\left(-\frac{d_{\ss\T}^{2}}{4D_{0}t}\right),
\end{align}
where $D_{0}$ is the diffusion coefficient of type $A_0$ molecules in $\frac{\m^{2}}{\s}$ and $d_{\ss\T}$ is the distance between the TX and $\RX_k$ in $\m$. We denote $S_{\ob}^{\ss{\R_k}}(t)$ as the number of molecules observed within $V_{\ss\R_k}$ at time $t$ due to the emission of molecules from the TX's current and previous symbols. We label the value of the realization of $S_{\ob}^{\ss{\R_k}}(t_{\ss\R}(j,m))$ as $s_{j,m}^{\ss{\R_k}}$. $S_{\ob}^{\ss{\R_k}}(t)$ can be accurately approximated by a Poisson random variable (RV) with the mean given by
\begin{align}\label{observed molecular numbers R}
\bar{S}_{\ob}^{\ss{\R_k}}(t)
= &\;\sum\limits^{\lfloor{\frac{t}{T}+1}\rfloor}_{i=1}S_{0}W_{\ss\T}[i]P_{\ob}^{({\ss{\T},{\R_k}})}(t-(i-1)T).
\end{align}

The sum of $M_{\ss\RX}$ samples by $\RX_k$ in the $j$th symbol interval, $S_{\ob}^{\ss{\R_k}}[j] = \sum_{m=1}^{M_{\ss\RX}}S_{\ob}^{\ss{\R_k}}(t_{\ss\R}(j,m))$, is also a Poisson RV whose mean is given by $\bar{S}_{\ob}^{\ss{\R_k}}[j] = \sum_{m=1}^{M_{\ss\RX}}\bar{S}_{\ob}^{\ss{\R_k}}(t_{\ss\R}(j,m))$. We label the value of the realization of $S_{\ob}^{\ss{\R_k}}[j]$ as $s_{j}^{\ss{\R_k}}$. 

\subsubsection{$\RX_k-\FC$ Link}\label{RX-FC link}
The RXs detect with an energy detector, i.e., $\RX_k$ compares $s_{j}^{\ss{\R_k}}$ with a \emph{constant} decision threshold at $\RX_k$, $\xi_{\ss\R_k}$, that is independent of $\textbf{W}_{\ss\T}^{j-1}$.
We denote $\hat{W}_{{\ss\R}_k}[j]$ as $\RX_k$'s binary decision on the $j$th transmitted symbol. We define $\hat{\textbf{W}}^{l}_{\ss\RX_k}=\left\{\hat{W}_{\ss\RX_k}[1],\ldots,\hat{W}_{\ss\RX_k}[l]\right\}$ as an $l$-length subsequence of the binary decisions at $\RX_k$. We denote $P_{\ob}^{({\ss\R_k,\ss\FC})}(t)$ as the probability of observing a given $A_1$ molecule, emitted from the center of $\RX_k$ at $t=0$, inside $V_{\ss\FC}$ at time $t$. Due to the relatively close proximity between the RXs and the FC, we find that \eqref{probability} cannot be used to evaluate $P_{\ob}^{({\ss\R_k,\ss\FC})}(t)$. Thus, we resort to \cite[Eq. (27)]{Noel2013} to evaluate $P_{\ob}^{(\ss\R_k,\ss\FC)}(t)$ as
\begin{align}\label{general prob}
P_{\ob}^{({\ss\R_k,\ss\FC})}(t)=&\;\frac{1}{2}\left[\erf\left(\tau_{1}\right)+\erf\left(\tau_{2}\right)\right]\nonumber\\
&-\frac{\sqrt{D_{{1}}t}}{d_{\ss\FC_k}\sqrt{\pi}}\left[\exp\left(-\tau_{1}^{2}\right)-\exp\left(-\tau_{2}^{2}\right)\right],
\end{align}
where $\tau_{1}=\frac{r_{\ss\FC}+d_{\ss\FC_k}}{2\sqrt{D_{{1}}t}}$,  $\tau_{2}=\frac{r_{\ss\FC}-d_{\ss\FC_k}}{2\sqrt{D_{{1}}t}}$, $D_{{1}}$ is the diffusion coefficient of type $A_{1}$ molecules in $\frac{\m^{2}}{\s}$, and $d_{\ss\FC_k}$ is the distance between $\RX_k$ and the $\FC$ in $\m$.

\section{ML Detection Design and Analysis}\label{sec:ML detection design}
In this section, we present \emph{symbol-by-symbol} ML detection design for the cooperative MC system. We denote $\hat{W}_{\ss\FC}[j]$ as the FC's decision on the $j$th symbol transmitted by the TX. We define $\hat{\textbf{W}}^{l}_{\ss\FC}=\left\{\hat{W}_{\ss\FC}[1],\ldots,\hat{W}_{\ss\FC}[l]\right\}$ as an $l$-length subsequence of the FC's decisions on the symbols transmitted by the TX. We note that ML detection in the $j$th symbol interval requires the knowledge of ISI due to the previously-transmitted symbols by the TX (and by all the RXs for noisy reporting with SD-ML). For practicality, the FC relies on local history to choose the current symbol, but for tractability we use the genie-aided history when deriving the system error probability. We denote $\hat{W}_{\ss\FC_k}[j]$ as the FC's estimated binary decision of $\RX_k$ on the $j$th transmitted symbol. We define $\hat{\textbf{W}}^{l}_{\ss\FC_k}=\left\{\hat{W}_{\ss\FC_k}[1],\ldots,\hat{W}_{\ss\FC_k}[l]\right\}$ as the FC's estimate of the first $l$ binary decisions by $\RX_k$. For \emph{given} local history at the FC, we formulate the general decision rule of ML detection on the $j$th symbol transmitted by the TX as
\begin{align}\label{ML rule}
\hat{W}_{\ss\FC}[j]=\underset{W_{\ss\T}[j]}{\text{argmax}}~ \mathcal{L}\left[j|\hat{\textbf{W}}_{\ss\FC}^{j-1}\right]
\end{align}
or
\begin{align}\label{ML rule,DF}
\hat{W}_{\ss\FC}[j]=\underset{W_{\ss\T}[j]}{\text{argmax}}~ \mathcal{L}\left[j|\hat{\textbf{W}}_{\ss\FC}^{j-1},\hat{\textbf{W}}^{j-1}_{\ss\FC_k}\right].
\end{align}

Eq. \eqref{ML rule} applies to F-ML and L-ML. Eq. \eqref{ML rule,DF} applies to SD-ML. For the sake of simplicity, we define $\mathcal{L}\left[j\right]\triangleq\mathcal{L}\left[j|\hat{\textbf{W}}_{\ss\FC}^{j-1}\right]$ for F-ML and L-ML and $\mathcal{L}\left[j\right]\triangleq\mathcal{L}\left[j|\hat{\textbf{W}}_{\ss\FC}^{j-1},\hat{\textbf{W}}^{j-1}_{\ss\FC_k}\right]$ for SD-ML.
In the remainder of this section, we present the detailed communication process and $\mathcal{L}\left[j\right]$ for F-ML, L-ML, and SD-ML. We denote $Q_{\ss\FC}[j]$ as the error probability of the cooperative MC system in the $j$th symbol interval for a TX sequence $\textbf{W}_{\ss\T}^{j-1}$. We derive $Q_{\ss\FC}[j]$ for L-ML to provide an analytical lower bound on the achievable error performance of the system in either the perfect or the noisy reporting scenario. By averaging $Q_{\ss\FC}[j]$ over all symbol intervals and all possible realizations of $\textbf{W}_{\ss\T}^{j-1}$, the average error probability of the cooperative MC system, $\overline{Q}_{\ss\FC}$, can be obtained.


\subsection{Perfect Reporting}
In the perfect reporting scenario, we assume that the FC has perfect knowledge of the observations at all RXs. Thus, we only consider one-phase transmission of each information symbol from the TX to the RXs. 

\subsubsection{ML detection}
Here, we present F-ML and L-ML.

\textbf{F-ML}: The FC separately assesses the likelihood of every sample by each RX and chooses the symbol $\hat{W}_{\ss\FC}[j]$ that is more likely, given the joint likelihood of $KM_{\ss\RX}$ individual observations at all RXs in the $j$th symbol interval. Since all individual observations are independent, $\mathcal{L}\left[j\right]$ is given by

\begin{align}\label{variant 1}
\mathcal{L}\left[j\right]=&\;\prod\limits^{K}_{k=1} \prod\limits^{M_{\ss\RX}}_{m=1}
\text{Pr}\left(S_{\ob}^{\ss{\R_k}}(t_{\ss\R}(j,m))=s_{j,m}^{\ss{\R_k}}|W_{\ss\T}[j],\hat{\textbf{W}}_{\ss\FC}^{j-1}\right).
\end{align}

\textbf{L-ML}: The FC adds each RX's $M_{\ss\RX}$ observations in the $j$th symbol interval, i.e., the FC applies an equal weight to every observation at each RX.
The FC chooses the symbol $\hat{W}_{\ss\FC}[j]$ that is more likely, given the joint likelihood of $K$ sums of $s_{j}^{\ss{\R_k}}$ in the $j$th symbol interval. Recalling our assumption that the $K$ RXs are independent, $\mathcal{L}\left[j\right]$ is given by
\begin{align}\label{variant 2}
\mathcal{L}\left[j\right]=\prod\limits^{K}_{k=1}\text{Pr}\left(S_{\ob}^{\ss{\R_k}}[j]=s_{j}^{\ss{\R_k}}|W_{\ss\T}[j],\hat{\textbf{W}}_{\ss\FC}^{j-1}\right).
\end{align}

It can be shown that \eqref{variant 1} and \eqref{variant 2} can be evaluated by the conditional probability mass function (PMF) of the Poisson RVs $S_{\ob}^{\ss{\R_k}}(t)$ and $S_{\ob}^{\ss{\R_k}}[j]$. The conditional mean of $S_{\ob}^{\ss{\R_k}}(t)$ given $\hat{\textbf{W}}_{\ss\FC}^{j-1}$ in \eqref{variant 1} is obtained by replacing $W_{\ss\T}[i]$ with $\hat{W}_{\ss\FC}[i]$ in \eqref{observed molecular numbers R}, where $i\in\{1,\ldots, j-1\}$. 


\subsubsection{Error Probability}
We now derive $Q_{\ss\FC}[j]$ for L-ML using \emph{genie-aided} history. To this end, we first define $\lambda_{\n,k}[j]$ as the expected ISI at $\RX_k$ in the $j$th symbol interval due to the previous TX symbols $\textbf{W}_{\ss\T}^{j-1}$, i.e.,
\begin{align}\label{noise, Va2}
\lambda_{\n,k}[j]
=\sum\limits^{j-1}_{i=1}S_{0}W_{\ss\T}[i]\sum\limits^{M_{\ss\RX}}_{m=1}P_{\ob}^{({\ss{\T},{\ss\R_k}})}\left(\left(j-i\right)T + m\Delta{t_{\ss\R}}\right).
\end{align}

We define $\lambda_{\s,k}[j]$ as the number of molecules at $\RX_k$ in the $j$th interval due to the current TX symbol $W_{\ss\T}[j]=1$, i.e.,
\begin{align}\label{signal, Va2}
\lambda_{\s,k}[j]
=&\;S_{0}\sum\limits^{M_{\ss\RX}}_{m=1}P_{\ob}^{({\ss{\T},{\ss\R_k}})}\left(m\Delta{t_{\ss\R}}\right).
\end{align}

When $\textbf{W}_{\ss\T}^{j-1} \neq \textbf{0}$, we have $\lambda_{\n,k}[j]>0$. When $\textbf{W}_{\ss\T}^{j-1} = \textbf{0}$, we have $\lambda_{\n,k}[j]=0$. Since we assume (via symmetry) that the RXs have independent and identically distributed observations, we have $\lambda_{\n,k}[j]= \lambda_{\n}[j]$ and $\lambda_{\s,k}[j]=\lambda_{\s}[j]$. For the sake of brevity, for L-ML, we define $\mathcal{L}\left[j|W_{\ss\T}[j]=1,\textbf{W}_{\ss\T}^{j-1}\right]\triangleq\mathcal{L}_1$ and $\mathcal{L}\left[j|W_{\ss\T}[j]=0,\textbf{W}_{\ss\T}^{j-1}\right]\triangleq\mathcal{L}_0$. Applying the conditional PMF of the Poisson RV $S_{\ob}^{\ss{\R_k}}[j]$ to \eqref{variant 2}, we write $\mathcal{L}_1$ and $\mathcal{L}_0$ as
\begin{align}\label{V2,L1}
\mathcal{L}_1=&\;\prod\limits^{K}_{k=1}\frac{\left(\lambda_\s[j]+\lambda_\n[j]\right)^{s_{j}^{\ss{\R_k}}}}{s_{j}^{\ss{\R_k}}!}
\exp\left(-\left(\lambda_\s[j]+\lambda_\n[j]\right)\right)
\end{align}
and
\begin{align}\label{V2,L0}
\mathcal{L}_0=&\;\prod\limits^{K}_{k=1}\frac{\left(\lambda_\n[j]\right)^{s_{j}^{\ss{\R_k}}}}{s_{j}^{\ss{\R_k}}!}\exp\left(-\lambda_\n[j]\right),
\end{align}
respectively. Based on \eqref{V2,L1} and \eqref{V2,L0}, we rewrite the general decision rule of L-ML in \eqref{ML rule} as the lower-complexity decision rule in the following theorem.
\begin{theorem}\label{theorem L-ML}
When $\lambda_\n[j]>0$, the decision rule of L-ML is to compare $s_{j}^{\ss{\R}}$ with an \emph{adaptive} threshold $\xi_{\ss\FC}^{\ad}[j]$, i.e.,
\begin{align}\label{variant 2 re1}
\hat{W}_{\ss\FC}[j]=
\begin{cases}
1,&\mbox{if $s_{j}^{\ss{\R}}\geq\xi_{\ss\FC}^{\ad}[j]$,}\\
0,&\mbox{$s_{j}^{\ss{\R}}<\xi_{\ss\FC}^{\ad}[j]$},
\end{cases}
\end{align}
where $s_{j}^{\ss{\R}} = \sum^{K}_{k=1}s_{j}^{\ss{\R_k}}$ and $\xi_{\ss\FC}^{\ad}[j] = \left\lfloor\frac{K\lambda_\s[j]}{\ln\left(\frac{\left(\lambda_\s[j]+\lambda_\n[j]\right)}{\lambda_\n[j]}\right)}\right\rceil$.
When $\lambda_\n[j]=0$, the decision rule of L-ML is to compare $s_{j}^{\ss{\R}}$ with $0$ and it is obtained by replacing $\geq$, $<$, and $\xi_{\ss\FC}^{\ad}[j]$ with $>$, $=$, and $0$ in \eqref{variant 2 re1}, respectively.
\end{theorem}
\begin{IEEEproof}
Applying \eqref{V2,L1} and \eqref{V2,L0} to \eqref{ML rule}, we write the decision rule of L-ML as
\begin{align}\label{variant 2 ML2}
\frac{\left(\lambda_\s[j]+\lambda_\n[j]\right)^{s_{j}^{\ss{\R}}}}{\exp\left(K\left(\lambda_\s[j]+\lambda_\n[j]\right)\right)}
\overset{\hat{W}_{\ss\FC}[j]=1}{\underset{\hat{W}_{\ss\FC}[j]=0} \gtreqless\;}\frac{\left(\lambda_\n[j]\right)^{s_{j}^{\ss{\R}}}}{\exp\left(K\lambda_\n[j]\right)},
\end{align}
When $\lambda_\n[j]>0$, we can further write \eqref{variant 2 ML2} as
\begin{align}\label{variant 2 ML3}
\left(\frac{\lambda_\s[j]+\lambda_\n[j]}{\lambda_\n[j]}\right)^{s_{j}^{\ss{\R}}}\overset{\hat{W}_{\ss\FC}[j]=1}{\underset{\hat{W}_{\ss\FC}[j]=0} \gtreqless\;}\exp\left(K\lambda_\s[j]\right).
\end{align}

We rearrange \eqref{variant 2 ML3} and then obtain \eqref{variant 2 re1}. If $\lambda_\n[j]=0$ and $s_{j}^{\ss{\R}}=0$, then we can write \eqref{variant 2 ML2} as
\begin{align}\label{variant 2 ML4}
\exp\left(-K\lambda_\s[j]\right){<}1,
\end{align}
where the decision at the FC is always $\hat{W}_{\ss\FC}[j]=0$. If $\lambda_\n[j]=0$ and $s_{j}^{\ss{\R}}>0$, then we can write \eqref{variant 2 ML2} as
\begin{align}\label{variant 2 ML5}
\left(\lambda_\s[j]\right)^{s_{j}^{\ss{\R}}}\exp\left(-K\lambda_\s[j]\right)>0,
\end{align}
where the decision at the FC is always $\hat{W}_{\ss\FC}[j]=1$. This completes the proof.
\end{IEEEproof}

Based on Theorem \ref{theorem L-ML}, when $\textbf{W}_{\ss\T}^{j-1} \neq \textbf{0}$, we evaluate $Q_{\ss\FC}[j]$ for L-ML as
\begin{align}\label{perfect probability1}
Q_{\ss\FC}[j] = &\;\left(1-P_{1}\right)\textrm{Pr}\left(S_{\ob}^{\ss{\R}}[j] \geq\xi_{\ss\FC}^{\ad}[j]|W_{\ss\T}[j]=0,\textbf{W}_{\ss\T}^{j-1}\right)\nonumber\\
&+P_1\textrm{Pr}\left(S_{\ob}^{\ss{\R}}[j] <\xi_{\ss\FC}^{\ad}[j]|W_{\ss\T}[j]=1,\textbf{W}_{\ss\T}^{j-1}\right),
\end{align}
where $S_{\ob}^{\ss{\R}}[j]= \sum^{K}_{k=1}S_{\ob}^{\ss{\R_k}}[j]$. When $\textbf{W}_{\ss\T}^{j-1} = \textbf{0}$, we obtain $Q_{\ss\FC}[j]$ for L-ML by replacing $\geq$, $<$, and $\xi_{\ss\FC}^{\ad}[j]$ with $>$, $=$, and $0$ in \eqref{perfect probability1}, respectively.

%
%

\subsection{Noisy Reporting}
In the noisy reporting scenario, the transmission of each information symbol from the TX to the FC via the RXs is completed in two phases. The first phase of noisy reporting is analogous to the one-phase transmission of perfect reporting. For the second phase, we consider SD-ML between the RXs and the FC over a diffusive channel.

\textbf{SD-ML}: Each $\RX_k$ transmits type $A_{1}$ molecules to report $\hat{W}_{{\ss\R}_k}[j]$ to the FC. Similar to the TX, each RX uses ON/OFF keying to report its decision to the FC and the RX releases $S_1$ molecules of type $A_1$ to convey information symbol ``1''. The FC receives type $A_1$ molecules over all $K$ $\RX_{k}-\FC$ links and takes $M_{\ss\FC}$ samples of type $A_1$ molecules in every reporting interval.
The FC adds $M_{\ss\FC}$ observations for all $\RX_k-\FC$ links in the $j$th symbol interval\footnote{To decrease the computational complexity at the FC, the FC assesses the likelihood of the \emph{sum} of $M_{\ss\FC}$ observations, which is analogous to L-ML.}. We denote ${S}_{\ob}^{{\ss\FC},\D}[j]$ as the total number of $A_1$ molecules observed at the FC in the $j$th symbol interval, due to both current and previous emissions of molecules by all RXs. Since the TX and $\RX_k$ use the same modulation method and the $\TX-\RX_k$ and $\RX_k-\FC$ links are both diffusive, ${S}_{\ob}^{{\ss\FC},\D}[j]$ can be accurately approximated by a Poisson RV. We denote ${\bar{S}}_{\ob}^{\ss\FC,\D}[j]$ as the mean of ${S}_{\ob}^{\ss\FC,\D}[j]$. We label the value of the realization of ${S}_{\ob}^{\ss\FC,\D}[j]$ as $s_{j}^{\ss\FC}$. The FC chooses the symbol $\hat{W}_{\ss\FC}[j]$ that is more likely, given the likelihood of $s_{j}^{\ss\FC}$ in the $j$th interval. To present $\mathcal{L}\left[j\right]$ for SD-ML, we first define $\hat{\mathcal{W}}^{{\ss\RX}}_j=\{\hat{W}_{\ss\R_1}[j]\ldots\hat{W}_{\ss\R_K}[j]\}$ as the set of decisions at all RXs in the $j$th symbol interval. We then define a set $\mathcal{R}$ which includes all possible realizations of $\hat{\mathcal{W}}^{{\ss\RX}}_j$ and the cardinality of set $\mathcal{R}$ is $2^K$. 
We derive $\mathcal{L}\left[j\right]$ as
\begin{align}\label{SD-ML}
\mathcal{L}\left[j\right]=
&\;\sum_{\hat{\mathcal{W}}^{{\ss\RX}}_j\in\mathcal{R}}\textrm{Pr}\left(\hat{\mathcal{W}}^{{\ss\RX}}_j|W_{\ss\T}[j],\hat{\textbf{W}}_{\ss\FC}^{j-1}\right)\nonumber\\
&\times\textrm{Pr}\left({S}_{\ob}^{\ss\FC,\D}[j]=s_{j}^{\ss\FC}|\hat{\mathcal{W}}^{{\ss\RX}}_j,\hat{\textbf{W}}_{\ss\FC_1}^{j-1},\ldots\hat{\textbf{W}}_{\ss\FC_K}^{j-1}\right),
\end{align}
where we clarify that we need to consider every realization of $\hat{\mathcal{W}}^{{\ss\RX}}_j$ and the corresponding probability to lead to $s_{j}^{\ss\FC}$. However, in \eqref{SD-ML}, it is hard for the FC to obtain $\hat{\textbf{W}}_{\ss\FC_k}^{j-1}$ when all RXs transmit type $A_1$ molecules to the FC in SD-ML. Fortunately, due to the symmetric topology, $\hat{\textbf{W}}_{\ss\FC_k}^{j-1}$ is not precisely required for calculating the conditional PMF of ${S}_{\ob}^{\ss\FC,\D}[j]$ in \eqref{SD-ML}; only the number of RXs that transmitted symbol ``1'' in each previous symbol interval is needed. We first define $\mathcal{Z}$ as the set where the elements are the possible number of RXs that transmit symbol ``1'' in the each symbol interval, i.e., $\mathcal{Z}=\{0,1,\ldots,K\}$. We denote $\hat{Z}[j]$ as the FC's estimate of the number of RXs that transmit symbol ``1'' in the $j$th symbol interval. We define $\hat{\textbf{Z}}^{l}=\left\{\hat{Z}[1],\ldots,\hat{Z}[l]\right\}$ as the FC's estimate of the number of RXs transmitting the symbol ``1'' in the first $l$ symbol intervals. Hence, for the evaluation of the likelihood in all future intervals, i.e., $\mathcal{L}\left[j+1\right],\ldots,\mathcal{L}\left[L\right]$, the FC also chooses $\hat{Z}[j]$ in the $j$th interval from the set $\mathcal{Z}$ that is most likely, given the likelihood of $s_{j}^{\ss\FC}$ in the $j$th interval. $\hat{Z}[j]$ is obtained by
\begin{align}\label{ML rule_RX_SD}
\hat{Z}[j]=\underset{Z}{\text{argmax}}~ \textrm{Pr}\left({S}_{\ob}^{\ss\FC,\D}[j]=s_{j}^{\ss\FC}|Z\in\mathcal{Z},\hat{\textbf{Z}}^{j-1}\right).
\end{align}

Also, since the cooperative MC system has a symmetric topology, the RXs have independent and \emph{identically} distributed observations. This leads to $\textrm{Pr}\left(\hat{W}_{\ss\R_k}[j]=1|W_{\ss\T}[j],\hat{\textbf{W}}_{\ss\FC}^{j-1}\right)=\textrm{Pr}\left(\hat{W}_{\ss\R}[j]=1|W_{\ss\T}[j],\hat{\textbf{W}}_{\ss\FC}^{j-1}\right)\triangleq\Theta_j$. Using $\hat{\textbf{Z}}^{j-1}$ and the notation $\Theta_j$, we rewrite \eqref{SD-ML} as
\begin{align}\label{SD-ML1}
\mathcal{L}\left[j\right]=
&\;\sum\limits^{K}_{Z=0}\binom{K}{Z}{\Theta_j}^{Z}\left(1-\Theta_j\right)^{K-Z}\nonumber\\
&\times\textrm{Pr}\left({S}_{\ob}^{\ss\FC,\D}[j]=s_{j}^{\ss\FC}|Z\in\mathcal{Z},\hat{\textbf{Z}}^{j-1}\right),
\end{align}
which can be evaluated by applying the conditional cumulative distribution function (CDF) of the Poisson RV $S_{\ob}^{\ss{\R_k}}[j]$ and the conditional PMF of the Poisson RV ${S}_{\ob}^{{\ss\FC},\D}[j]$. Finally, we derive the conditional mean of ${S}_{\ob}^{\ss\FC,\D}[j]$ given $Z$ and $\hat{\textbf{Z}}^{j-1}$. We first write ${\bar{S}}_{\ob}^{\ss\FC,\D}[j]$ as
\begin{align}\label{observed molecular numbers R1_SD1}
{\bar{S}}_{\ob}^{\ss\FC,\D}[j]
=&\;S_1\sum\limits^{j-1}_{i=1}\sum\limits^{K}_{k=1}\hat{W}_{\ss\FC_k}[i]\nonumber\\
&\times\sum\limits^{M_{\ss\FC}}_{\tilde{m}=1}P_{\ob}^{({\ss\R_k,\ss\FC})}\left(\left(j-i\right)T + \tilde{m}\Delta{t_{\ss\FC}}\right)\nonumber\\
&+S_1\sum\limits^{K}_{k=1}\hat{W}_{\ss\RX_k}[j]\sum\limits^{M_{\ss\FC}}_{\tilde{m}=1}P_{\ob}^{({\ss\R_k,\ss\FC})}\left(\tilde{m}\Delta{t_{\ss\FC}}\right),
\end{align}

Since our topology is symmetric, we then rewrite \eqref{observed molecular numbers R1_SD1} as
\begin{align}\label{observed molecular numbers R1_SD2}
{\bar{S}}_{\ob}^{\ss\FC,\D}[j]
=&\;S_1\sum\limits^{j-1}_{i=1}\hat{Z}[i]\sum\limits^{M_{\ss\FC}}_{\tilde{m}=1}P_{\ob}^{({\ss\R_k,\ss\FC})}\left(\left(j-i\right)T + \tilde{m}\Delta{t_{\ss\FC}}\right)\nonumber\\
&+S_1Z\sum\limits^{M_{\ss\FC}}_{\tilde{m}=1}P_{\ob}^{({\ss\R_k,\ss\FC})}\left(\tilde{m}\Delta{t_{\ss\FC}}\right).
\end{align}

\section{Numerical Results and Simulations}\label{sec:Numerical}
In this section, we present numerical and simulation results to examine the error performance of ML detection for the cooperative MC system. We simulate using a particle-based method considered in \cite{Andrew2004}, where the precise locations of all individual molecules are known. Since we model the RXs and the FC as passive observers, in our simulation all molecules persist indefinitely once they are released. Unless otherwise stated, we consider the environmental parameters in Table~\ref{tab:table1}. Since we consider a symmetric topology, we assume the same decision threshold at all RXs such that $\xi_{\ss\RX_k} = \xi_{\ss\RX}, \forall k$.
\begin{table}[!t]
\renewcommand{\arraystretch}{1.2}
\centering
\caption{Environmental Parameters}\label{tab:table1}\vspace{-1mm}
\begin{tabular}{c||c|c}
\hline
\bfseries Parameter &  \bfseries Symbol&  \bfseries Value \\
\hline\hline
Radius of RXs& $r_{\ss\R_k}$ & $0.2\,{\mu}\metre$\\\hline
Radius of FC & $r_{\ss\FC}$ & $0.2\,{\mu}\metre$ \\\hline
Time step at RXs & $\Delta{t_{\ss\R}}$ & $50\,{\mu}\s$\\\hline
Time step at FC & $\Delta{t_{\ss\FC}}$ & $30\,{\mu}\s$ \\\hline
Number of samples by RXs& $M_{\ss\RX}$ & 5 \\\hline
Number of samples by FC& $M_{\ss\FC}$ & 10 \\\hline
Transmission time interval & $t_{\trans}$ & $0.5\,{\m}\s$\\\hline
Report time interval & $t_{\report}$ & $0.3\,{\m}\s$\\\hline
Bit interval time& $T$ & $0.8\,{\m}\s$\\\hline
Diffusion coefficient & $D_0=D_{1}$ & $5\times10^{-9}{\m^{2}}/{\s}$\\\hline
Length of symbol sequence & $L$ & $20$ \\\hline
Probability of binary 1 & $P_1$ & $0.5$ \\\hline
Number of cooperative RXs & $K$ & $4$ \\

\hline
\end{tabular}
\end{table}

\begin{table}[!t]
\renewcommand{\arraystretch}{1.2}
\centering
\caption{Devices' Location}\label{tab:coordinates1}\vspace{-1mm}
\begin{tabular}{c||c|c|c}
\hline
\bfseries Devices & \bfseries X-axis [${\mu}\metre$] & \bfseries Y-axis [${\mu}\metre$] & \bfseries Z-axis [${\mu}\metre$]\\\hline\hline
$\TX$   & $0$ & $0$ & $0$\\\hline
$\RX_1$ & $2$ & $0.6$ & $0$\\\hline
$\RX_2$ & $2$ & $-0.6$ & $0$ \\\hline
$\RX_3$ & $2$ & $-0.3$ & $0.5196$ \\\hline
$\RX_4$ & $2$ & $-0.3$ & $-0.519$ \\\hline
$\FC$   & $2$ & $0$ & $0$ \\
\hline
\end{tabular}
\end{table}

In the following, we assume that the TX releases $S_{0} = 5000$ molecules for symbol ``1'' and for SD-ML, each RX releases $S_1=500$ molecules to report a decision of ``1''. We consider a system that consists of at most four RXs. The specific locations of the TX, RXs, and FC are listed in Table~\ref{tab:coordinates1}. Furthermore, the simulated error probabilities are averaged over independent transmissions of $5\times10^4$ randomly generated TX symbol sequences.

In order to provide insights in terms of the trade-offs among the error performance, the knowledge of previous symbols, and computational complexity, we compare the error performance of the proposed ML detection for a cooperative MC system with that of the following alternatives:
\begin{enumerate}
\item The majority rule which we considered in \cite{GC2016} and shows the best error performance among all hard fusion rules. In the majority rule, the behavior of each RX is similar
to that in SD-ML, but each RX reports its decision to the FC using a unique type of molecule and the FC combines the received decisions at all RXs and declares a decision of ``1'' on the TX's symbol when it receives at least $\lceil K/2\rceil$ decisions of ``1''.
\item A simple soft fusion variant. In~\cite{GC2016} we proposed this variant to provide a simple lower bound on the error performance of the cooperative system with hard fusion rules. In this variant, the FC adds all RXs' $KM_{\ss\RX}$ samples in the $j$th symbol interval, and then makes a decision $\hat{W}_{\ss\FC}[j]$ by comparing this sum $s_{j}^{\ss{\R}}$ with a \emph{constant} threshold $\xi_{\ss\FC}$, that is independent of $\textbf{W}_{\ss\T}^{j-1}$.
\end{enumerate}
For these two variants, we consider the same parameters as for the ML detection variants listed in Table~\ref{tab:table1} and described above. This ensures that our comparisons are fair.

In the following figures, we clarify that the value of $\overline{Q}_{\ss\FC}^{\ast}$ is the minimum $\overline{Q}_{\ss\FC}$ by numerically optimizing the corresponding constant decision thresholds of SD-ML, the majority rule, and the simple soft fusion variant. In the following figures, for each ML detection variant, we plot the error probability using the local history and genie-aided history. 

In the following figures, we observe that for all ML detection variants, the error performance using local history has a very small degradation from that using the genie-aided history. This demonstrates the effectiveness of our proposed method for the FC to estimate the symbols previously transmitted by the TX (and by all the RXs for SD-ML). We also observe that the simulations precisely match the analytical results, thereby validating our analytical results, in particular the analytical error performance of L-ML using genie-aided history.


\begin{figure}[!t]
\centering
\includegraphics[height=2.35in,width=0.95\columnwidth]{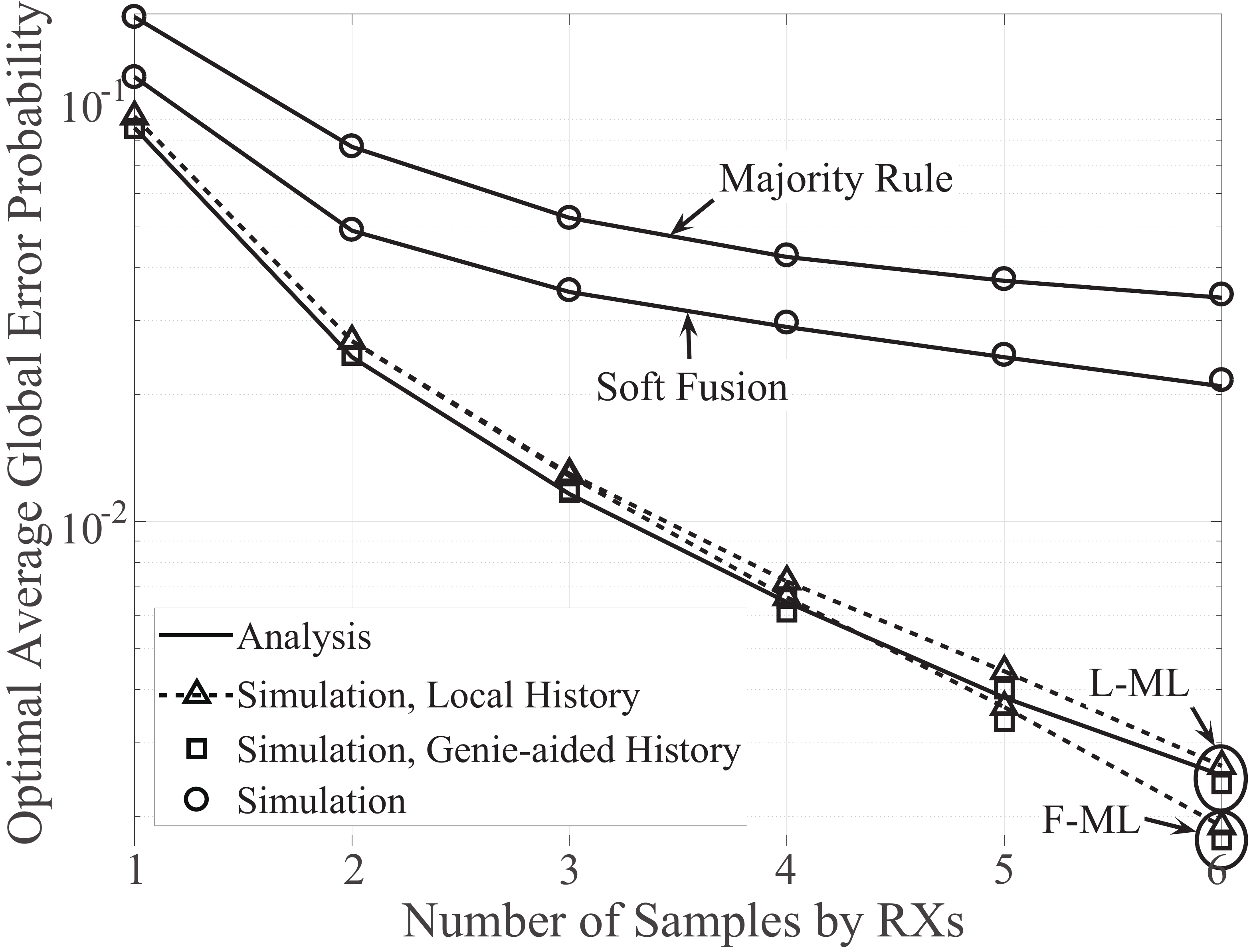}
\caption{Optimal Average global error probability $\overline{Q}_{\ss\FC}^{\ast}$ of different variants versus the number of samples by RXs $M_{\ss\RX}$ in the perfect reporting scenario.}
\label{fig:perfect}
\end{figure}

In Fig.~\ref{fig:perfect}, we consider the perfect reporting scenario. We plot the optimal average global error probability versus the number of samples by RXs for F-ML, L-ML, the majority rule, and the soft fusion variant. In this figure, the report time interval is fixed at $t_{\trans}=0.5\,{\m}\s$ as in Table~\ref{tab:table1} and the time step at the RXs for each $M_{\ss\RX}$ is $\Delta{t_{\ss\RX}}=t_{\trans}/2M_{\ss\RX}$. We first observe that F-ML and L-ML achieve a significant error performance improvement over the majority rule and the soft fusion variant. This demonstrates the advantage of ML detection for the cooperative MC system, even though the ML detection is applied on a symbol-by-symbol basis. Second, we observe that for the majority rule, the error performance gain for the soft fusion variant is minimal relative to that achieved with ML detection. This observation is not surprising since the decision rule of L-ML is comparing the sum $s_{j}^{\ss{\R}}$ with the \emph{adaptive} threshold $\xi_{\ss\FC}^{\ad}[j]$, while the soft fusion variant compares this sum with the \emph{constant} threshold $\xi_{\ss\FC}$. Third, we observe that F-ML outperforms L-ML. This is because the likelihood of every sample by each RX is considered separately in F-ML, which entails higher computational complexity compared to L-ML, whereas only the sum of all samples by each RX are considered in L-ML. Finally, we see that the system error performance improves as $M_{\ss\RX}$ increases.

\begin{figure}[!t]
\centering
\includegraphics[height=2.35in,width=0.95\columnwidth]{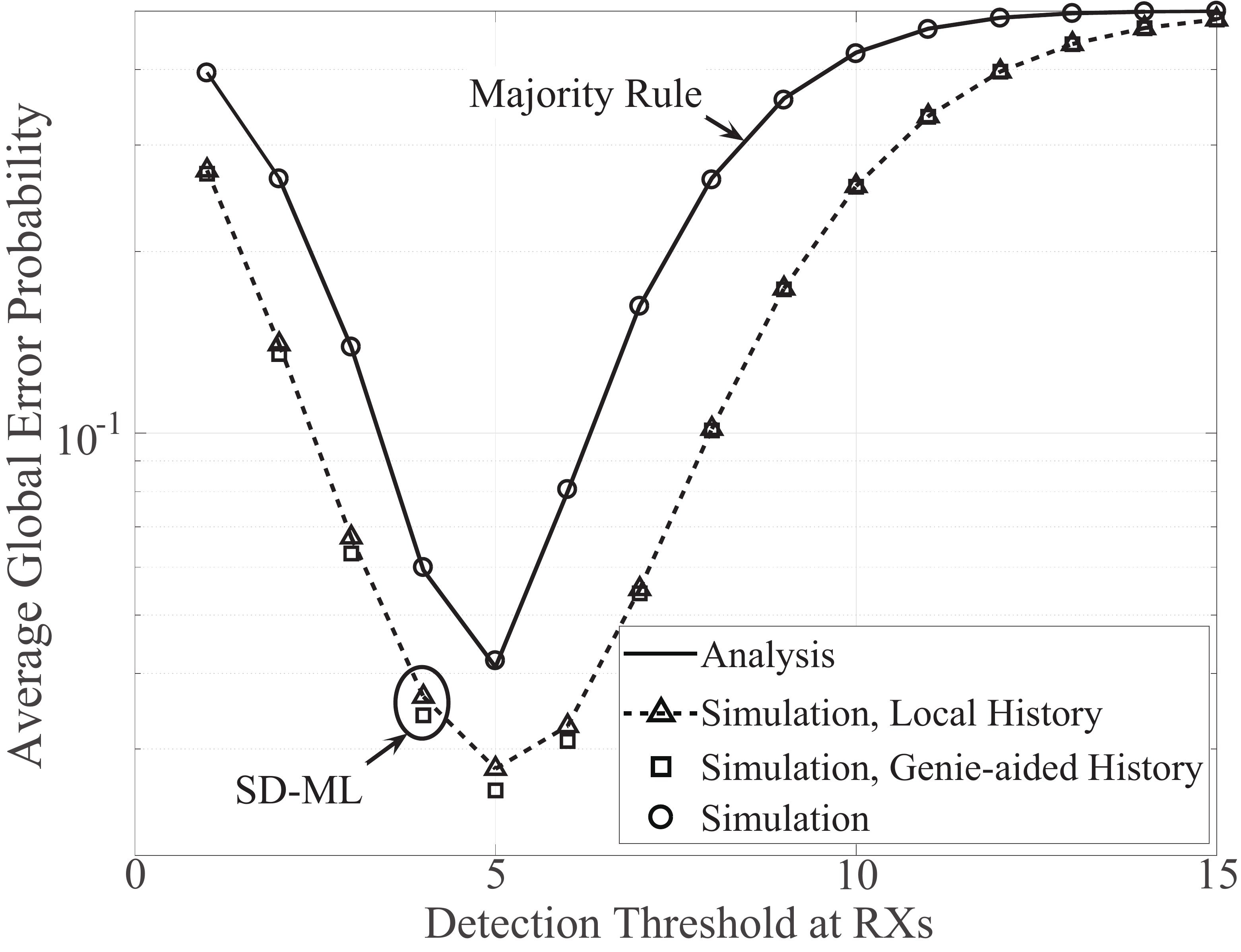}
\caption{Average global error probability $\overline{Q}_{\ss\FC}$ of different variants versus the decision threshold at RXs, $\xi_{\ss\R}$ in the noisy reporting scenario.}
\label{fig:SD-threshold}
\end{figure}

In Fig. \ref{fig:SD-threshold}, we consider the noisy reporting scenario. We plot average global error probability versus the decision threshold at RXs for SD-ML and the majority rule. In this figure, we consider $\xi_{\ss\FC}=4$ for the majority rule, since $\xi_{\ss\FC}$ is this value when the thresholds at the RXs and the FC are jointly optimized. This ensures the fairness of our comparison. We observe that SD-ML outperforms the majority rule, i.e., SD-ML provides lower bounds on the error probability for the majority rule. We also observe that the majority rule only suffers a $34\%$ error performance degradation compared to SD-ML using local history at their corresponding optimal RX detection thresholds. This demonstrates the good performance of the majority rule, relative to the SD-ML variant. 

\begin{figure}[!t]
\centering
\includegraphics[height=2.4in,width=0.95\columnwidth]{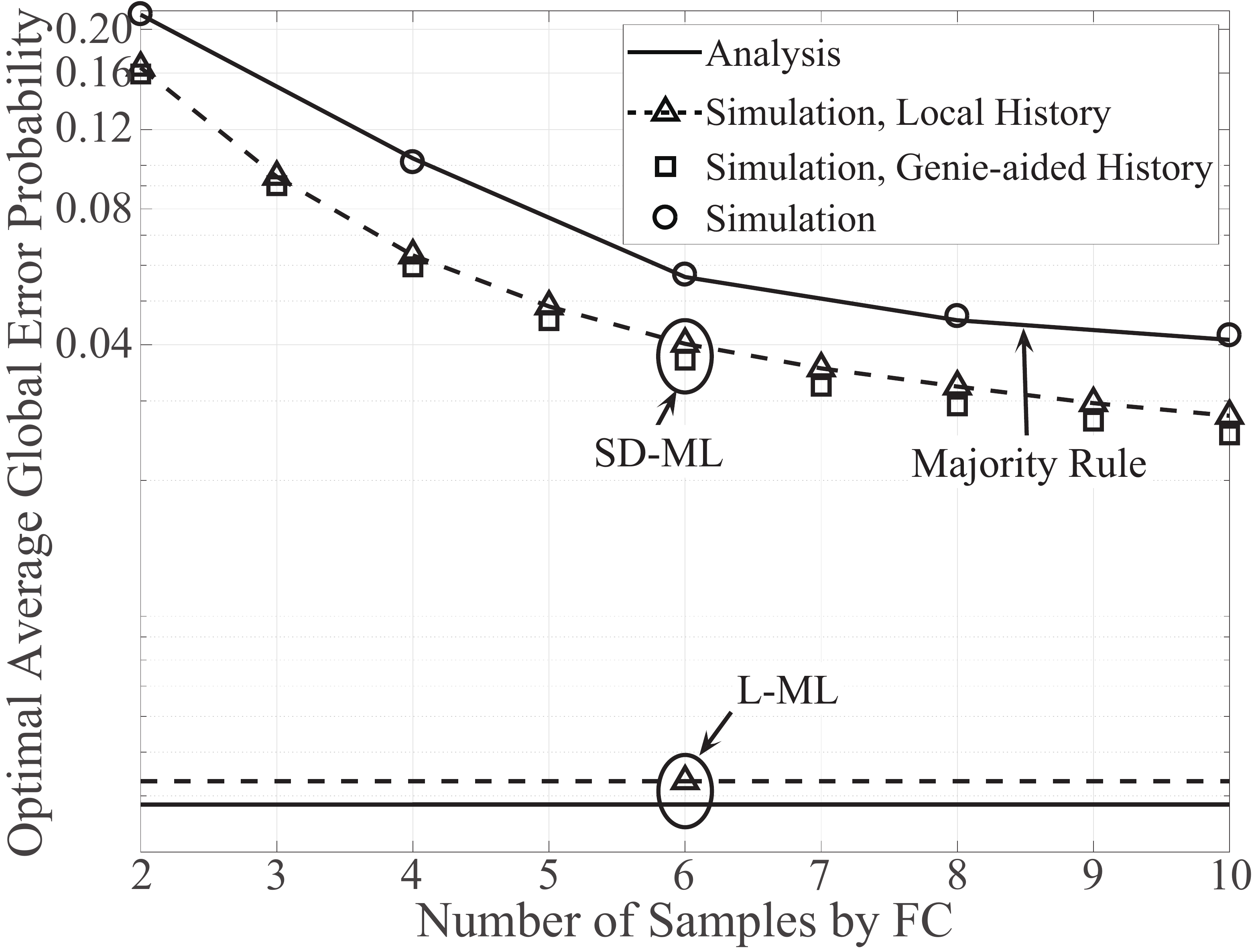}
\caption{Optimal Average global error probability $\overline{Q}_{\ss\FC}^{\ast}$ of different variants versus the number of samples by FC $M_{\ss\FC}$ in the noisy reporting scenario. We emphasize that $\overline{Q}_{\ss\FC}$ is independent of $M_{\ss\FC}$ for L-ML.}
\label{fig:SD-M_FC}
\end{figure}

In Fig. \ref{fig:SD-M_FC}, we plot the optimal average global error probability versus the number of samples by the FC for SD-ML and the majority rule in the noisy reporting scenario and L-ML in the perfect reporting scenario. We first observe that SD-ML outperforms the majority rule. Second, we observe that the majority rule achieves a moderately worse error performance compared to SD-ML. For example, when $M_{\ss\FC}=6$, the majority rule only suffers a moderate $30\%$ error performance degradation compared to SD-ML using local history. This observation is consistent with our observations in Fig. \ref{fig:SD-threshold} and it again demonstrates the good performance of the majority rule, relative to the SD-ML variant. We note that in the majority rule, each RX releases a unique type of molecule, but the FC requires relatively low complexity. However, in SD-ML, all RXs release a single type of molecule, but the FC requires relatively high complexity. Thus, SD-ML is more suitable for an environment with high complexity at the FC and a limited number of molecule types, whereas the majority rule is more suitable for an environment with sufficient types of molecules but limited computational capability at the FC.  
Importantly, we observe that L-ML has a significant improvement over SD-ML and the majority rule, i.e., L-ML provides a lower bound on the achievable error performance of the system in either the perfect or the noisy reporting scenario. Given that F-ML outperforms L-ML, as observed in Fig.~\ref{fig:perfect}, we clarify that F-ML achieves the best error performance and SD-ML achieves worst error performance among three ML detection variants. It is noted that SD-ML is the most realistic and feasible variant among them, since we consider a diffusive channel between the RXs and the FC in SD-ML. Finally, we see that the system error performance improves as $M_{\ss\FC}$ increases.

\vspace{0.2cm}

\section{Conclusions}\label{sec:con}\vspace{0.1cm}
In this paper, we presented for the first time symbol-by-symbol ML detection for the cooperative diffusion-based MC system with multiple communication phases. This approach enables us to determine lower bounds on the error performance of the simpler cooperative variants. We presented three ML detection variants with different constraints on the knowledge available at the FC, i.e., F-ML, L-ML, and SD-ML. We derived a closed-form expression for the system error probability for L-ML and corroborated the accuracy of this expression using simulation results. We demonstrated the good performance of the majority rule relative to SD-ML, e.g., the majority rule suffers a $34\%$ error performance degradation compared to SD-ML in Fig. \ref{fig:SD-threshold}. Our results revealed the trade-off between the system error performance and the knowledge available at the FC. In our future work, we will investigate ML detection variants with other constraints on the knowledge available at the FC, e.g., amplify-and-forward relaying at the RXs.



\end{document}